\documentclass[10pt,twocolumn]{IEEEtran}
\usepackage{lipsum}
\usepackage{flushend}
\usepackage{amsbsy,amsmath,amsthm,amsfonts,amssymb,amsbsy,subfigure,verbatim,etoolbox}
\usepackage{bm,cite,graphicx,psfrag,pstricks,tikz,times,url,verbatim}
\usetikzlibrary{shapes,snakes,calendar,matrix,backgrounds,folding}
\usepackage[english]{babel}
\usetikzlibrary{arrows,automata}
\usetikzlibrary{calc} 
\usetikzlibrary{decorations.pathmorphing} 

\newtoggle{Arxiv}
\toggletrue{Arxiv}

\newcommand{\bi}{\begin{itemize}}
\newcommand{\ei}{\end{itemize}}
\newcommand{\ben}{\begin{enumerate}}
\newcommand{\een}{\end{enumerate}}

\newcommand{\bc}{\begin{cases}}
\newcommand{\ec}{\end{cases}}
\newcommand{\bd}{\begin{description}}
\newcommand{\ed}{\end{description}}

\newcommand{\be}{\begin{equation}}
\newcommand{\ee}{\end{equation}}
\newcommand{\bea}{\begin{eqnarray}}
\newcommand{\eea}{\end{eqnarray}}

\newtheorem{thm}{Theorem}

\newtheorem{ass}{Assumption}

\newtheorem{algo}{Algorithm}

\theoremstyle{plain}
\newtheorem{remark}{Remark}

\newcommand{\nm}[1]{\textcolor{red}{\textbf{[NM: #1]}}}

\graphicspath{%
     {./Figures/}
}

\begin{document}

\title{Energy-Based Adaptive Multiple Access in LPWAN IoT Systems with Energy Harvesting}
\author{Nicolo Michelusi and Marco Levorato
\thanks{N. Michelusi is with the School of Electrical and Computer Engineering, Purdue University. \emph{email}: michelus@purdue.edu.}
\thanks{M. Levorato is with the School of Information and Computer Sciences, University of California, Irvine. \emph{email}: levorato@uci.edu.}
}
\maketitle

\begin{abstract}
This paper develops a control framework for a network of energy harvesting nodes connected to a Base Station (BS) over a multiple access channel. The objective is to adapt their transmission strategy to the state of the network, including the energy available to the individual nodes. In order to reduce the complexity of control, an optimization framework is proposed where
\emph{energy storage dynamics} are replaced by \emph{dynamic average} power constraints induced by the time correlated energy supply, thus enabling 
lightweight and flexible network control. Specifically, the BS adapts the packet transmission probability of 
the "active" nodes (those currently under a favorable energy harvesting state) so as to maximize the average long-term throughput, under these
\emph{dynamic average} power constraints. The resulting policy takes the form of the packet transmission probability as a function of the energy harvesting state and number of active nodes. The structure of the throughput-optimal genie-aided policy, in which the number of active nodes is known non-causally at the BS, is proved. 
Inspired by the genie-aided policy, a Bayesian estimation approach is presented to address the case where the BS estimates the number of active nodes based on the observed network transmission pattern.
It is shown that the proposed scheme outperforms by $20$\% a scheme
in which the nodes operate based on local state information only,
and performs well even when energy storage dynamics are taken into account.
\end{abstract}

\section{Introduction}
Important technological trends within the Internet of Things (IoT) domain, such as Smart City and Urban IoT systems~\cite{IoT}, push for the development of network solutions providing long range communication capabilities to mobile devices distributed over large geographical areas. As a consequence, wireless cellular networks may play a key role toward the practical deployment of such systems.

However, recent cellular network standards are not designed to support machine-to-machine and computational services such as those that will characterize future large-scale IoT systems. On the one hand, in most cases IoT devices and services generate sporadic and low-intensity traffic. On the other hand,  the potentially huge number of IoT devices interconnected through a single base station (BS) would raise new issues related to the signaling and control traffic, which may become the bottleneck of the system. Low-Power Wide Area Networks (LPWAN), and its 
LoRaWAN\textsuperscript{TM} specification \cite{lorawanspec}, were proposed to meet these characteristics and requirements~\cite{vangelista2015long,centenaro2015long}, and provide connectivity between short range (\emph{e.g.}, Bluetooth) and long-range cellular communications. These networks exploit unlicensed frequency bands to create star topologies directly connected to a unique collector node, generally referred to as the \emph{gateway}. The architecture of these networks is especially designed to provide wide area coverage and ensure connectivity to a large number of low-power devices.

Another key enabler of the IoT is energy harvesting \cite{Paradiso}, which enables long-term and self-sustaining sensing and communication operations \cite{Gunduz}. Among
the many energy harvesting technologies, \emph{e.g.}, vibration, light, and thermal energy extraction, wireless energy harvesting~\cite{kamalinejad2015wireless} is one of the most promising solutions due to its simplicity, ease of implementation, and wide availability. However, the limited energy input rate of these harvesting technologies especially in mobile environments is only suitable for simple applications with low on-device sensing, processing and communication requirements. LPWAN technologies are a perfect match to this class of applications. However, sensible design of channel access strategies with minimal 
coordination and control overhead
is necessary to efficiently use the scarce energy resources.

In this paper, an optimization framework is proposed which makes the channel access strategy of the connected nodes aware of their \emph{harvesting state}, that is, the input rate of energy to the batteries. Consistently with LPWAN technologies, a simple centralized architecture is considered, where the central coordinator, that is, the BS, sporadically controls the transmission probability of the wireless nodes. For this scenario, under the assumption of binary Markovian energy harvesting state, the structure of the throughput-optimal transmission policy is derived under energy constraints and full network state knowledge (genie-aided). 
Inspired by the optimal genie-aided policy, 
a Bayesian estimation framework is presented for the case where the coordinator needs to estimate the harvesting state of the nodes in order to make network control.

LPWAN are characterized by a potentially large number of devices accessing a unique BS. 
This requirement poses a severe challenge in network design and optimization, in that the state space of the network
grows exponentially with the network size as $\mathcal S^N$,
where $\mathcal S$ is the state space of a single node.
This is especially cumbersome
in LPWAN systems with energy harvesting.
In fact, in these systems, the state  of each node $s\in\mathcal S$ specifies the state of charge of the rechargeable battery, which may even be difficult to estimate \cite{MichelusiSOC,valentini2016optimal,valentini2016aging},
as well as the state of the ambient energy source (\emph{e.g.}, "high" or "low" as in \cite{MichelusiCORR}).
Therefore, network adaptation should be  based on the state and dynamics of the energy storage element of each node, which can become
unmanageable even for small networks \cite{DelTesta}.
Thus, \emph{there is a need to develop complexity reduction techniques for the design, analysis and optimization of these networks.}
In this paper, we propose to reduce the network state space and  enable lightweight and flexible network control
by replacing energy storage dynamics with \emph{dynamic average} power constraints induced by the time correlated energy supply.
This approach removes the need to perform adaptation based on the current state of charge of each device. Instead, adaptation is 
done solely based on the state of the ambient energy harvesting process.

The design of energy harvesting networks has seen huge interest in the research community \cite{Gatzianas,Varan,Ulukus}. The problem of random access, similar to this paper, has been considered in \cite{MichelusiRAND}, for the case with i.i.d. energy harvesting, and assuming the nodes operated based on local state information only.
Energy management policies under time-correlated energy harvesting have been studied in \cite{MichelusiCORR} for a single node.
In this paper, we extend these results to \emph{multiple} access networks under \emph{time-correlated} energy harvesting,
and provide a form of \emph{network control} (as opposed to local control).

Numerical results show that the
energy harvesting states provide a natural mechanism of partial network coordination: the devices
 can tune their transmission parameters based on the estimated network state, and reduce the detrimental effect of collisions. The proposed strategy outperforms by 20\% a fully decentralized scheme where nodes make decisions solely based on their local energy harvesting state.
Our proposed approximation is shown to perform well even when battery dynamics are taken into account.

This paper is organized as follows. In Sec. \ref{sec:sysmo}, we present the system model and, in Sec. \ref{analysis}, the analysis; we provide numerical results in Sec. \ref{sec:numres}, and concluding remarks in Sec.~\ref{sec:concl}.
\iftoggle{Arxiv}{%
The main proofs and algorithms are provided in the Appendix.
}{
Due to space constraints, the proofs are provided in \cite{MicheLevo}.
}
\section{System Model}
\label{sec:sysmo}

Consider a network of $N$ energy harvesting (EH) nodes, indexed by $n=1,2,\dots, N$,
communicating over a shared channel to a gateway.
Time is slotted with slot duration $T$, and transmissions are synchronous.

{\bf Energy harvesting model:}
Each node harvests ambient energy. We model the harvested energy as i.i.d. across nodes.
In each node, the harvested energy is characterized by a Markovian state with two states $\{L,H\}$, where
$L$ denotes the "low" EH state, and $H$ the "high" EH state.
We let $\lambda_S,S\in\{L,H\}$ be the average power harvested in state $S\in\{L,H\}$, with $\lambda_H>\lambda_L\geq 0$.\footnote{Herein, we do not assume any specific
distribution of harvested power in the "high" and "low" states.}
Thus, in state $S$, a node receives, on average, $\lambda_S T$ Joules of energy in one slot.
 We denote the state of the $n$th node in slot $k$ as 
$S_{n,k}\in\{L,H\}$. The  transition probability from $L$ to $H$ is denoted as $p_H$,
and that from $H$ to $L$ as $p_L$,
where $1-p_H-p_L>0$ (positive memory).
 Hence, at steady state,
\begin{align}
\label{EHsteady}
&\pi_S\triangleq\mathbb P(S_{n,k}=S)=\frac{p_S}{p_H+p_L},\ \forall S\in\{H,L\}.
\end{align}

{\bf Battery dynamics:}
Each node has an internal energy storage element (either rechargeable battery or super-capacitor)
of capacity $e_{\max}$ [Joules]
 to store the harvested ambient energy. 
Denote the internal state of node  $n$ at the beginning of slot $k$ as $E_{n,k}$. This state evolves according to the dynamics
\begin{align}
\label{ehoperation}
E_{n,k+1}=\min\left\{E_{n,k}-C_{n,k}+A_{n,k},e_{\max}\right\},
\end{align}
where $0\leq C_{n,k}\leq E_{n,k}$ is the energy consumed by the node in slot $k$,  and $A_{n,k}\geq 0$ is the energy harvested in slot $k$.
The dynamics of $\{E_{n,k},k\geq 0\}$ over the network $n\in\{1,2,\dots,N\}$ introduce a severe design challenge,
since $(E_{1,k},E_{2,k},\dots,E_{N,k})\in [0,e_{\max}]^N$ becomes part of the network state.
The exponential growth of the state space with $N$
 challenges the practical optimization and analysis of communication and networking protocols. 

In order to reduce the state space of the system and thus enable lightweight and flexible network control, we
note that the time correlation in the harvested ambient energy, coupled with battery dynamics,
approximately induce
  a state dependent average power constraint
  \begin{align}
\label{powconstraint}
\mathbb E[P_{n,k}|S_{n,k}=S]\leq \lambda_S,\ \forall S\in\{H,L\},
\end{align}
where $P_{n,k}=C_{n,k}/T$ is the average power consumed. In fact, if this power constraint is exceeded, the battery discharges 
leading to energy outage.

Motivated by this observation,
in this paper we neglect the battery dynamics (\ref{ehoperation}), and 
we replace them with the average power constraints (\ref{powconstraint}) in the "high" and "low" EH states.
This approach significantly reduces the complexity of network control, since the network state
$(E_{1,k},E_{2,k},\dots,E_{N,k})\in [0,e_{\max}]^N$
and its dynamics given by (\ref{ehoperation})
 need not be taken into account.
 The difficulty with this approach is that the average power constraint varies dynamically and randomly with the energy harvesting
 state, as opposed to battery-powered networks, where the power constraint is fixed.
 This limitation is overcome by network control, developed in Sec. \ref{analysis}.

{\bf Communication model:}
Each node is backlogged with data to transmit. Transmissions occur probabilistically for each node according to a random access scheme,
with transmission power $P_{tx}>0$.
We let $q_{n,k}$ be the transmission probability of node $n$ in slot $k$,
as specified in Sec. \ref{analysis}.

Each node randomly chooses one of $B{>}0$  orthogonal channels for transmission.
 We assume a collision model, \emph{i.e.},
 the transmission succeeds if and only if one node transmits on a given channel.
For instance, in a CDMA based system \cite{Padovani}, $B$ corresponds to the number of 
orthogonal spreading sequences, chosen randomly from each transmitting device.
If two devices select the same spreading sequence, then a collision occurs and the transmission fails.
In contrast, if they select mutually orthogonal spreading sequences, then they can suppress their mutual interference
and the transmission succeeds. 
 
Based on this model, the instantaneous expected throughput,
function of the vector of transmission probabilities across the network,
$\mathbf q_k=(q_{1,k},q_{2,k},\dots,q_{N,k})$, is given by
\begin{align}
\label{reward}
r(\mathbf q_k)\triangleq \sum_{n=1}^Nq_{n,k}\prod_{n\neq m}\left(1-\frac{q_{m,k}}{B}\right).
\end{align}
In fact, node $n$ transmits in channel $i\in\{1,2,\dots, B\}$ with probability $q_{n,k}/B$.
The transmission succeeds if none of the other nodes transmit on the same channel, with probability $\prod_{n\neq m}\left(1-\frac{q_{m,k}}{B}\right)$.
The expression (\ref{reward}) is finally obtained by adding together the individual throughputs in each channel,
and for each node.

{\bf Performance metric and optimization problem:}
For this transmission model, the power constraints in
(\ref{powconstraint}) induce a 
constraint on the transmission probabilities given by
\begin{align}
\label{constraints}
&\bar Q_S(\mu)\triangleq\mathbb E_\mu[q_{n,k}|S_{n,k}=S]\leq \frac{\lambda_S}{P_{tx}},\ \forall S\in\{H,L\}.
\end{align}
Since $q_{n,k}{\leq}1$, the constraint becomes inactive when $\lambda_S{\geq}P_{tx}$.
We define the average long-term network throughput as
\begin{align}
\label{Rmu}
\bar R(\mu)=\lim_{T\to\infty}\mathbb E\left[\frac{1}{T}\sum_{k=0}^{T-1}r(\mathbf q_k)\right].
\end{align}
Both (\ref{constraints}) and (\ref{Rmu})
 are functions of some policy $\mu$, which governs the selection of transmission probabilities by each device,
depending on the information available at the central or local controller.
The goal is to determine $\mu^*$ so as to maximize the network throughput, \emph{i.e.},
\begin{align}
\label{sdfhdfgh}
\mu^*{=}\arg\max\bar R(\mu),
\text{ s.t. }\bar Q_S(\mu)\leq\frac{\lambda_S}{P_{tx}},\ \forall S\in\{H,L\}.
\end{align}

In the next section, we
address the optimization problem (\ref{sdfhdfgh}) by
 considering three scenarios differing in the amount of state information available at the local or central controller.
In this paper, we focus on the special case $\lambda_L{=}0$ (no energy harvested in the "low" EH state, so that $q_{n,k}{=}0$ when $S_{n,k}{=}L$)
and one channel $B{=}1$. 
We leave the more general case $\lambda_L{\geq}0$ and $B{\geq}1$ for future investigations.
\section{Analysis of adaptive multiple-access policies}
\label{analysis}
In this section, we design transmission policies for three different scenarios:
\begin{itemize}
\item \emph{Local EH state}, where each node has only local knowledge of its EH state $S_{n,k}$ (Sec. \ref{sec1});
\item \emph{Genie-aided}, where each node knows the number of
"active" nodes (those in the "high" EH state) (Sec. \ref{sec2});
\item \emph{Bayesian},  where the gateway infers the number of active nodes
based on the observed network operation; for this case, we will leverage the "genie-aided" policy to design a low-complexity policy applicable to this scenario of more practical interest (Sec. \ref{sec3}).
\end{itemize}
\subsection{Local EH state}
\label{sec1}
In this case, $q_{n,k}$ is a function of $S_{n,k}$ only.
We thus define the policy $q_{n,k}=\mu_{S_{n,k}}$,\footnote{We assume that the policy does not depend on $n$ or $k$, in order to simplify the design.}
where $\mu_{H}$ and $\mu_L$ are the transmission probabilities in the "high" and "low" EH states, respectively.
Thus, (\ref{constraints}) becomes
\begin{equation}
\label{constraints2}
\mu_H\leq \min\left\{1,\frac{\lambda_H}{P_{tx}}\right\},~~~~\mu_L=0,
\end{equation}
since $q_{n,k}=0$ for nodes in the "low" EH state.
At steady state,
the EH states are independent across the network, yielding
\begin{align}
\label{rewx}
\bar R_\mu=\mathbb E_\mu[r(\mathbf q_k)]=N\bar q\left(1-\bar q\right)^{N-1},
\end{align}
where we have defined the average long-term transmission probability for each node,
\begin{align}
\label{barq}
\bar q\triangleq\pi_H\mu_H+\pi_L\mu_L=\pi_H\mu_H.
\end{align}
Note from (\ref{constraints2}) that 
\begin{align}
\label{qmax}
\bar q\leq\pi_H\min\left\{1,\frac{\lambda_H}{P_{tx}}\right\}\triangleq q_{\max}.
\end{align}
By maximizing $\bar R_\mu$ in  (\ref{rewx}) over $0\leq\bar q\leq q_{\max}$, we obtain
\begin{align}
\bar q^*=\min\left\{q_{\max},\frac{1}{N}\right\},
\end{align}
yielding, from (\ref{barq}) and (\ref{qmax}),
\begin{align}
\mu_H^*=
\frac{\bar q^*}{\pi_H}
=
\min\left\{1,\frac{\lambda_H}{P_{tx}},\frac{1}{N\pi_H}\right\}.
\end{align}
\subsection{Genie-aided}
\label{sec2}
In the genie-aided case, node $n$ knows $S_{n,k}$ and the number of active nodes, denoted as $N_{H,k}\triangleq \sum_{m=1}^N\chi(S_{m,k}=H)$ at time $k$, where $\chi(\cdot)$ is the indicator function. Thus, 
$q_{n,k}$ is adapted based on $(S_{n,k},N_{H,k})$, according to the policy
\begin{align}
q_{n,k}=\mu_{S_{n,k}}(N_{H,k}).
\end{align}
Since $q_{n,k}{=}0$ in the "low" EH state, we have $\mu_{L}(m){=}0,\forall m\in\{0,1,\dots,N\}$.
At steady state,
the number of active nodes, node $n$ excluded,
 $N_{H,k}{-}\chi(S_{n,k}{=}H)$, is 
a binomial random variable with parameter $\pi_H$ and $N-1$ trials.
Thus, 
the average transmission probability in the "high" EH state is given by
\begin{align}
\label{P}
&\bar Q_H(\mu)
=\sum_{m=0}^{N-1}\mathbb P_\mu(N_{H,k}=m+1|S_{n,k}=H)\mu_H(m+1)
\nonumber
\\&
=
\sum_{m=0}^{N-1}\left(\!\!\!\begin{array}{c}N-1\\m\end{array}\!\!\!\right)\pi_H^m\pi_L^{N-1-m}\mu_H(m+1).
\end{align}
Similarly, the network throughput is given by
\begin{align}
\label{R}
&\!\!\bar R(\mu)
{=}\!\!
\sum_{m=1}^{N}\left(\!\!\begin{array}{c}N\\m\end{array}\!\!\right)\pi_H^m\pi_L^{N-m}
m\mu_H(m)\left(1{-}\mu_H(m)\right)^{m-1}\!,\!\!
\end{align}
since $N_{H,k}$ is binomial with $N$ trials and parameter $\pi_H$,
and each of the active nodes transmit with probability $\mu_H(N_{H,k})$.
The optimization problem thus becomes
\begin{align}
\mu_H^*(\cdot)=&\arg\max_\mu\bar R(\mu),
\text{ s.t. }\bar Q_H(\mu)\leq\frac{\lambda_H}{P_{tx}}.\nonumber
\end{align}

Theorem \ref{thm1} provides  the structure of $\mu_H^*(\cdot)$.
We let $\lambda_{H,\max}\triangleq\frac{P_{tx}}{N\pi_H}\left(1-\pi_L^N\right)$.
\begin{thm}
\label{thm1}
If
$\lambda_H\leq P_{tx}\pi_L^{N-1}$,
then 
\begin{equation}
\label{structure1}
\mu_H^*(1)=\frac{\lambda_H}{P_{tx}\pi_L^{N-1}},~~~
\mu_H^*(m)=0,\ \forall m>1.
\end{equation}
Otherwise, if 
$P_{tx}\pi_L^{N-1}{<}\lambda_H{<}\lambda_{H,\max}
$, then
\begin{align}
\mu_H^*(1)=1,
\ 
\mu_H^*(m),\ \forall m>1,
\end{align}
where $\mu_H^*(m),m{>}1$  is the unique $\mu_H(m){\in}(0,m^{-1})$ such that
 \begin{align}
  \label{mona2}
  \left(1-\mu_H(m)\right)^{m-2}\left(1-m\mu_H(m)\right)=\phi,\ \forall m>1,
\end{align}
and $\phi\in(0,1)$ is 
the unique value such that (\ref{P}) is satisfied with equality.
Finally, if 
$\lambda_H\geq\lambda_{H,\max}$,
then
\begin{align}
\label{case3}
\mu_H^*(m)=\frac{1}{m},\ \forall m\geq 1.
\end{align}
\end{thm}
\iftoggle{Arxiv}{%
\begin{proof}
See Appendix \ref{proofofthm1}.
\end{proof}
}{
}


According to Theorem \ref{thm1}, when $\lambda_H$ is small ($\leq P_{tx}\pi_L^{N-1}$),
transmissions are allowed only when a unique node is active ($N_{H,k}=1$).
In fact, allowing multiple nodes to transmit (when $N_{H,k}>1$) would incur performance degradation due to collisions.
On the other hand, for larger $\lambda_H$,
there is an energy surplus that can be used to allocate transmissions to multiple nodes also when $N_{H,k}>1$.
If, further, $\lambda_H<\lambda_{H,\max}$, the transmission probability constraint  (\ref{P}) is satisfied with equality.
However, when $\lambda_H\geq\lambda_{H,\max}$, the  constraint  (\ref{P}) becomes loose. 
This is because, with $N_{H,k}=m$ active nodes, the instantaneous expected throughput
$m\mu_H(m)(1{-}\mu_H(m))^{m-1}$
is maximized by $\mu_H^*(m)=\frac{1}{m}$. Transmitting with probability larger than $1/m$ would incur throughput degradation and higher energy cost.
Thus, there is no benefit in using the surplus of energy available.

In the previous theorem, 
when $P_{tx}\pi_L^{N-1}{<}\lambda_H{<}\lambda_{H,\max}$,
$\phi$ should be determined so that (\ref{P}) is satisfied with equality, with $\mu_H(m)$ given by the solution of (\ref{mona2}).
In order to solve this numerically, note that  the left hand side of (\ref{mona2}) is a strictly decreasing function of $\mu_H(m)\in(0,1/m)$.
Hence, the solution of (\ref{mona2})
can be determined via  the bisection method \cite{bisection}, and  is a decreasing function of $\phi$. From this it also follows that $\bar Q_H(\mu)$ is a decreasing function of $\phi$.
Thus, $\phi$ can be found numerically using the bisection method \cite{bisection}.
\iftoggle{Arxiv}{%
We refer the interested reader to the proof of Theorem \ref{thm1} in Appendix \ref{proofofthm1}.
Additionally, the bisection algorithms are provided in Appendix \ref{app_algo}.
}{
We refer the interested reader to \cite{MicheLevo} for details on these algorithms.
}
\subsection{Bayesian scheme}
\label{sec3}
In this case, the gateway observes the sequence $\{t_k,k\geq 0\}$ of
 the number of nodes that attempted transmission in slot $k$. This information becomes available at the gateway at the end of slot $k$, and, in practice, can be inferred by monitoring the interference level over the channel.
 We assume that the identity of these nodes is not known.
Based on $\mathbf t_k{=}(t_0,t_1,\dots,t_{k-1})$ available at the beginning of slot $k$, the gateway computes a posterior probability distribution (belief) over 
the number of active nodes $N_{H,k}$. Denote such belief as $\beta_k$ \emph{i.e.},
\begin{align}
\label{belief}
\beta_k(m)\triangleq\mathbb P(N_{H,k}=m|\mathbf t_k,\mu^{(k)}),m=0,1,\dots,N,
\end{align}
where $\mu^{(k)}=(\mu_{H,0},\mu_{H,1},\dots,\mu_{H,k-1})$ is the vector of access probabilities used by the active nodes up to slot $k$. 
Given $\beta_k$, the gateway selects the transmission probability $\mu_k$ for the active nodes in slot $k$ and broadcasts this control information to the whole network.

Then, $t_{k}$  is observed and the new belief becomes
\begin{align}\nonumber
\label{betaupdate}
&\beta_{k+1}(m)=\mathbb P(N_{H,k+1}=m|\mathbf t_k,\mu^{(k)},t_{k},\mu_k)
\\&
=
\frac{
\sum_{m^\prime=t_k}^N\beta_{k}(m^\prime)\mathbb P(t_{k}|\mu_k,N_{H,k}=m^\prime)
\mathbb P_N(m|m^\prime)
}{
\sum_{m^\prime=t_k}^N\beta_{k}(m^\prime)\mathbb P(t_{k}|\mu_k,N_{H,k}=m^\prime)
},
\end{align}
where we have defined $\mathbb P_N(m|m^\prime){\triangleq}\mathbb P(N_{H,k+1}{=}m|N_{H,k}{=}m^\prime)$.
Above,
 $t_k$ is a binomial random variable with parameter $\mu_k$ and $N_{H,k}$ trials,
and thus $\mathbb P(t_{k}|\mu_k,N_{H,k}{=}m^\prime)$ is given by
\begin{align}
\mathbb P(t_{k}|\mu_k,N_{H,k}=m^\prime)=
\left(\begin{array}{c}m^\prime\\t_k\end{array}\right)\mu_k^{t_k}(1-\mu_k)^{m^\prime-t_k}.
\end{align}
Additionally,
 $N_{H,k+1}{=}N_{H,k}{-}x{+}y$, where $x{\leq}N_{H,k}$
is the number of nodes (out of $N_{H,k}$ nodes) that switch from the "high" to the "low" EH state, and $y{=}x{+}N_{H,k+1}{-}N_{H,k}$ is the number of nodes (out of $N{-}N_{H,k}$ nodes)
that become active,
 so that $\mathbb P_N(m|m^\prime)$
 is given by
 \begin{align}
& \mathbb P_N(m|m^\prime)
 \nonumber
 =
 \sum_{x=(m^\prime-m)^+}^{\min\{m^\prime,N-m\}}
\left(\begin{array}{c}m^\prime\\x\end{array}\right)\left(\begin{array}{c}N-m^\prime\\x+m-m^\prime\end{array}\right)
\\&
\times
p_L^x(1-p_L)^{m^\prime-x}
p_H^{x+m-m^\prime}(1-p_H)^{N-m-x}.
 \end{align}
 Note that $\mathbb P_N(m|m^\prime)$ is independent of $\mu_k$.
 Therefore, it can be computed only once at  initialization of the system, and updated when the EH conditions change.
 
 Since in this scenario information on  $N_{H,k}$ is only partially available, the optimization of the
 transmission probability $\mu_k$ as a function of the belief $\beta_k$ can be expressed as a Partially Observable Markov Decision Process \cite{Sondik}.
 This optimization has high complexity due to the high-dimensional belief space. 
 In this paper, in order to reduce the complexity,
 we choose $\mu_k$ so that, given $\beta_k$, the expected network power consumption is the same as  in the genie-aided case, \emph{i.e.},
\begin{align}
\label{P1}
P_{tx}\mu_{k}\sum_{m=1}^N\beta_{k}(m)m
=
P_{tx}\sum_{m=1}^N\beta_{k}(m)m\mu_H^*(m),
\end{align}
yielding
\begin{align}
\label{dfghcv}
\mu_{k}
=
\frac{\sum_{m=1}^N\beta_{k}(m)m\mu_H^*(m)}{\sum_{m=1}^N\beta_{k}(m)m}.
\end{align}

Under such $\mu_k$,
the instantaneous expected throughput for a given belief $\beta_k$ is given by
\begin{align}
\tilde r(\beta_k)=\sum_{m=1}^N\beta_{k}(m)m\mu_{k}\left(1-\mu_{k}\right)^{m-1}.
\end{align}

Note that any \emph{feasible} scheme with partial network state information should satisfy the power constraints (\ref{constraints}).
With $\mu_k$ given by (\ref{dfghcv}), this is guaranteed by the following theorem.
\begin{thm}
\label{thm2}
Under the policy $\mu_k$ in (\ref{dfghcv}), the average power consumption in the "high" EH state 
is the same as that under the genie-aided scheme.
\end{thm}
\iftoggle{Arxiv}{%
\begin{proof}
See Appendix \ref{proofofthm2}.
\end{proof}
}{
}
\section{Numerical Results}
\label{sec:numres}
We provide simulation results for a system with parameters:
$N{=}20$ nodes; transition probabilities
$p_H{=}4\times 10^{-3}$ and
$p_L{=}20\times 10^{-3}$; normalized transmission power $P_{tx}=1$.
The harvesting power in the "high" EH state, $\lambda_H$, is varied in $[0,\lambda_{H,\max}]$.

\begin{figure}
    \centering
\includegraphics[width=0.8\linewidth,trim={0 5 40 20},clip]{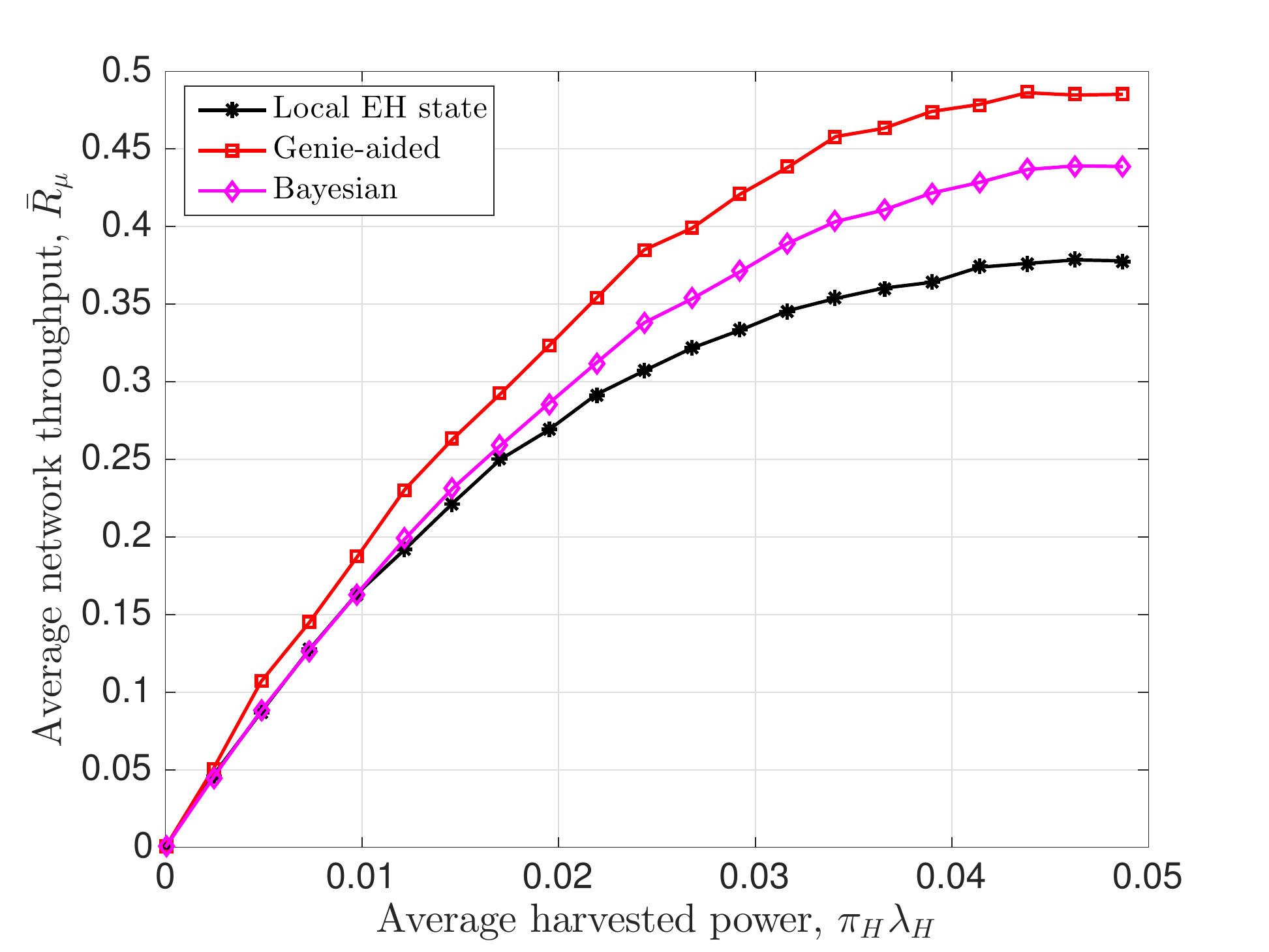}
\caption{Network throughput vs average harvested energy per node.}
\label{figexlabel}
\end{figure}

In Fig. \ref{figexlabel}, we plot the curve of the network throughput $\bar R_{\mu}$ versus the average harvested power per node $\pi_H\lambda_H$,
 obtained by varying $\lambda_H\in[0,\lambda_{H,\max}]$.
As expected, \emph{Local EH state} performs the worst, due to the lack of coordination among nodes.
In contrast, \emph{Genie-aided} performs the best: the "high" and "low" EH states provide a natural mechanism of partial coordination for the nodes,
which can tune their transmission parameters based on the number of active nodes.
Finally, \emph{Bayesian} exhibits intermediate performance,
due to the imperfect knowledge on the number of active nodes.

\begin{figure}
    \centering
\includegraphics[width=0.8\linewidth,trim={0 5 40 20},clip]{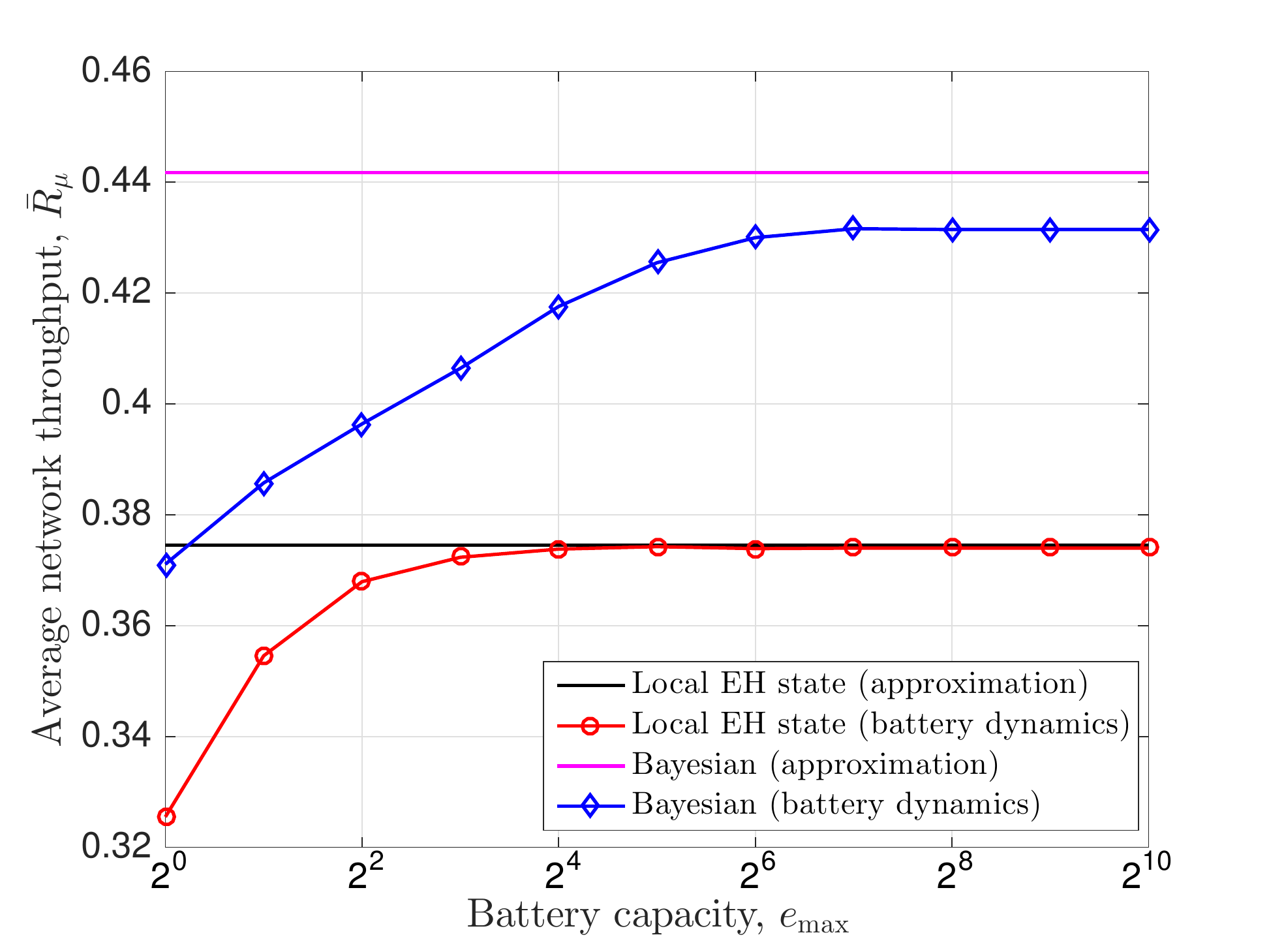}
\caption{Network throughput vs battery capacity.
"Approximation" refers to our proposed solution which neglects battery dynamics;
"battery dynamics" is an evaluation which takes into account these dynamics.
}
\label{fig2}
\iftoggle{Arxiv}{%
}{
}
\end{figure}

In Fig. \ref{fig2}, we evaluate via simulation the quality of the approximations introduced by 
replacing the battery dynamics (\ref{ehoperation}) with dynamic power constraints (\ref{powconstraint}).
We define an energy quantum as the quantity $P_{tx}T$, corresponding to the energy required to transmit over one slot.
We assume that the EH  process in  the "high" EH state is Bernoulli distributed, \emph{i.e.}, one energy quantum 
is received with probability $\lambda_H/P_{tx}$, otherwise no energy is received. Each transmission consumes one energy quantum.
For this evaluation, we let $\lambda_H=\lambda_{H,\max}$.
Note that the
 performance under "battery dynamics" incurs a performance degradation with respect to
their corresponding "approximation".
This is a result of the fact that, when the battery dynamics are taken into account, 
\emph{energy outage} (empty battery)
and \emph{energy overflow} (full battery)
may occur.
 The degradation decreases for larger $e_{\max}$, since energy overflow becomes less significant.
Nevertheless, \emph{Bayesian} outperforms  \emph{Local EH state} by up to $20\%$, even when battery dynamics
are accounted for. Thus, the approximation developed in this paper reduces significantly the optimization complexity, while
preserving the goodness of the solutions.
Interestingly, for large battery capacity, \emph{Local EH state} evaluated  under  "battery dynamics" approaches its "approximation", whereas
 a gap remains in \emph{Bayesian}. This gap is due to the fact that,
 in \emph{Bayesian}, transmissions depend on the number of active nodes $N_{H,k}$,
 which exhibit temporal correlation.
 As a result, the transmission sequence of a node also exhibits temporal correlation.
This may cause larger fluctuations in the state of charge of the battery,
 and thus, more frequent energy outages and overflows. 
 This effect is less relevant in  \emph{Local EH state}, since transmissions are independent of $N_{H,k}$.
\section{Conclusions}
\label{sec:concl}
In this paper, we have considered 
the design of adaptive multiple access policies 
for LPWAN energy harvesting IoT systems.
In order to reduce the complexity of network control, we have proposed an optimization framework 
which replaces energy storage dynamics with \emph{dynamic average} power constraints induced by the time correlated energy supply.
We have derived 
the structure of the throughput-optimal genie-aided transmission policy and, based on it, we have proposed a Bayesian estimation approach to address
the more practical scenario where the number of "active" nodes needs to be estimated based  on the observed network transmission pattern. 
We have shown by simulation that the proposed scheme outperforms by 20\% a scheme
in which the nodes operate based on local state information only,
and performs well even when energy storage dynamics are taken into account.

\iftoggle{Arxiv}{%
\appendices
\section{Proof of Theorem \ref{thm1}}
\label{proofofthm1}
\begin{proof}
In order to prove this theorem, we present a general methodology.
Let $\mu$ be a policy such that there exist 
two distinct indices $m_1\neq m_2$, 
$m_1,m_2\in\{1,2,\dots, N\}$, such that $\mu_H(m_1)<1$ and $\mu_H(m_2)>0$.

Let $\mu^{(\delta)}$ be a new policy, parameterized by $\delta>0$, defined as
\begin{align}
\label{newpol}
\left\{\begin{array}{ll}
\mu^{(\delta)}_H(m)=\mu_H(m),&\forall m\notin\{m_1,m_2\},
\\
\mu^{(\delta)}_H(m_1)=\mu_H(m_1)+\delta,
\\
\mu^{(\delta)}_H(m_2)=\mu_H(m_2)-g(\delta),
\end{array}\right.
\end{align}
where $\delta>0$ is small enough to guarantee a feasible policy $\mu^{(\delta)}_H(m_1)\in[0,1]$, $\mu^{(\delta)}_H(m_2)\in[0,1]$, and $g(\delta)$
 is a function such that 
the average transmission probability under $\mu$ and $\mu^{(\delta)}$ is the same, \emph{i.e.},
\begin{align}
\label{dfgh}
\bar Q_H(\mu)=\bar Q_H(\mu^{(\delta)}).
\end{align}
Using (\ref{newpol}) in  (\ref{P}) and in (\ref{dfgh}), we obtain
\begin{align}
g(\delta)=\frac{(m_2-1)!(N-m_2)!}{(m_1-1)!(N-m_1)!}\left(\frac{p_L}{p_H}\right)^{m_2-m_1}\delta.
\end{align}
Note that $g(\delta)>0$, hence $\mu^{(\delta)}$ is obtained from $\mu$ by decreasing the transmission probability in state $(H,m_2)$,
and augmenting it in state $(H,m_1)$, in such a way as to preserve the average power consumption in the high EH state (see (\ref{dfgh})). This is doable, since  $\mu_H(m_1)<1$ and $\mu_H(m_2)>0$ by assumption.

Note that, if there exists $\delta>0$ (arbitrarily small) such that $\bar R(\mu^{(\delta)})>\bar R(\mu)$, then $\mu$ is strictly suboptimal and is outperformed by policy 
$\mu^{(\delta)}$, which thus achieves the same average power consumption as $\mu$, but strictly larger network reward. Equivalently, in the limit $\delta\to 0$,
we need to verify
\begin{align}
\bar R^\prime(\mu)\triangleq\lim_{\delta\to 0}\frac{\bar R(\mu_\delta)-\bar R(\mu)}{\delta}\lessgtr 0,
\end{align}
where $\bar R^\prime(\mu)$ is the derivative of $\bar R(\mu_\delta)$ with respect to $\delta$ at $0$.
If $\bar R^\prime(\mu)>0$, then there exists a sufficiently small $\delta>0$ such that $\bar R(\mu_\delta)>\bar R(\mu)$, hence $\mu$ is strictly suboptimal.
On the other hand,
if $\mu_H(m_1)\in(0,1)$ and $\mu_H(m_2)\in(0,1)$, in order for  $\mu$ to be optimal, it must necessarily satisfy $\bar R^\prime(\mu)=0$;
in fact, if $\bar R^\prime(\mu)>0$, then 
 there exists a sufficiently small $\delta>0$ such that $\bar R(\mu_\delta)>\bar R(\mu)$;
 in contrast, if $\bar R^\prime(\mu)<0$, then 
 there exists a sufficiently small and negative $\delta<0$ such that $\bar R(\mu_\delta)>\bar R(\mu)$.
 
 Using (\ref{R}), we can show that $\bar R^\prime(\mu)$ is given by
 \begin{align}
 \label{rprime}
 \nonumber
&\bar R^\prime(\mu)\\&{=}\!
\left(\!\!\!\begin{array}{c}N\\m_1\end{array}\!\!\!\right)
\!\pi_H^{m_1}\pi_L^{N-m_1}m_1\!\left(1{-}\mu_H(m_1)\right)^{m_1-2}\!\left(1{-}m_1\mu_H(m_1)\right)
\nonumber
\\&
{-}\!\left(\!\!\!\begin{array}{c}N\\m_2\end{array}\!\!\!\right)\!
\pi_H^{m_2}\pi_L^{N-m_2}m_2\!\left(1{-}\mu_H(m_2)\right)^{m_2-2}
\nonumber
\\&
\qquad\qquad\times\left(1{-}m_2\mu_H(m_2)\right)\!g^\prime(0)
\nonumber
 \\&
 \propto
\left(1-\mu_H(m_1)\right)^{m_1-2}\left(1-m_1\mu_H(m_1)\right)
\nonumber
\\&
\quad-\left(1-\mu_H(m_2)\right)^{m_2-2}\left(1-m_2\mu_H(m_2)\right)
\nonumber
 \\& 
 \triangleq f_{m_1,m_2}(\mu_H(m_1),\mu_H(m_2)),
 \end{align}
 where $\propto$ denotes proportionality up to a positive multiplicative factor, and $g^\prime(\delta)=\frac{\mathrm dg(\delta)}{\mathrm d\delta}$.
 Therefore, $\bar R^\prime(\mu)>0$ if and only if $f_{m_1,m_2}(\mu_H(m_1),\mu_H(m_2))>0$.
We use this framework to prove the structure of the optimal policy.

\paragraph{Case $\lambda_H\leq P_{tx}\pi_L^{N-1}$}
First, we prove by contradiction that $\mu^*_H(m)=0,\forall m>1$ as in (\ref{structure1}).
Thus, let $\mu$ be a policy that does not obey this requirement, \emph{i.e.},
 there exists $m_2>1$ such that  $\mu_H(m_2)>0$.

If $\mu_H(1)=1$, then $\bar Q_H(\mu)$ in (\ref{P}) satisfies
\begin{align}
&\bar Q_H(\mu)
=
\sum_{m=0}^{N-1}\left(\!\!\!\begin{array}{c}N-1\\m\end{array}\!\!\!\right)\pi_H^m\pi_L^{N-1-m}\mu_H(m+1)
\\&
=
\pi_L^{N-1}+\left(\!\!\!\begin{array}{c}N-1\\m_2-1\end{array}\!\!\!\right)\pi_H^{m_2-1}\pi_L^{N-m_2}\mu_H(m_2)
\nonumber
\\&
+\sum_{m=1,m\neq m_2-1}^{N-1}\left(\!\!\!\begin{array}{c}N-1\\m\end{array}\!\!\!\right)\pi_H^m\pi_L^{N-1-m}\mu_H(m+1)
>
\pi_L^{N-1},
\nonumber
\end{align}
since $\mu_H(m_2)>0$.
Since $\pi_L^{N-1}\geq \lambda_H/P_{tx}$ by assumption, we finally obtain
\begin{align}
&\bar Q_H(\mu)>
\pi_L^{N-1}\geq \frac{\lambda_H}{P_{tx}},
\end{align}
and thus the constraint (\ref{constraints}) is violated. Thus, necessarily $\mu_H(1)<1$ if $\mu_H(m_2)>0$.

Then, let $\mu$ be such that $\mu_H(1)<1$ and $\mu_H(m_2)>0$, for some $m_2>1$.
Note that, letting $m_1=1$, we have that $\mu_H(m_1)<1$ and $\mu_H(m_2)>0$. Therefore,
we can apply the framework developed in the preliminary part of the proof.
 We achieve a contradiction in the optimality of $\mu$ if $f_{m_1,m_2}(\mu_H(m_1),\mu_H(m_2))>0$.
Indeed, from (\ref{rprime}) we obtain
\begin{align}
\label{f1}
&f_{1,m_2}(\mu_H(1),\mu_H(m_2))
\nonumber
\\&
=1-(1-\mu_H(m_2))^{m_2-2}[1-m_2\mu_H(m_2)],
\end{align}
which is strictly positive for $\mu_H(m_2)>0$. 

We thus obtain a contradiction in the optimality of $\mu_H$. Necessarily the optimal policy is such that $\mu_H^*(m)=0,\forall m>1$.
We now optimize over $\mu_H^*(1)\in[0,1]$ to show (\ref{structure1}).
From (\ref{P}) and (\ref{R}), we have that 
\begin{align}
\label{x}
&\bar Q_H(\mu^*)=\pi_L^{N-1}\mu_H^*(1)\leq \frac{\lambda_H}{P_{tx}},
\\
&\bar R(\mu^*)=N\pi_H\pi_L^{N-1}\mu_H^*(1).
\label{y}
\end{align}
 $\bar R(\mu^*)$ is an increasing function of $\mu_H^*(1)$, and thus the maximum network reward 
 is achieved when $\bar Q_H(\mu^*)$ is attained with equality, \emph{i.e.},  $\mu_H^*(1)=\frac{\lambda_H}{P_{tx}\pi_L^{N-1}}$, thus proving the optimality of (\ref{structure1}).

\paragraph{Case $\lambda_H>P_{tx}\pi_L^{N-1}$}
In this case, we first prove that $\mu_H^*(1)=1$.
Let $\mu$ be a policy such that $\mu_H(1)<1$.
We have two cases:
\begin{itemize}
\item $\mu_H(m)=0,\forall m>1$: in this case, we can 
improve the network throughput and still satisfy the (\ref{constraints}) by setting $\mu_H(1)=1$, see  (\ref{x})-(\ref{y}).
\item $\exists m_2>1:\mu_H(m_2)>0$; as in (\ref{f1}), letting $m_1=1$,
we obtain $f_{1,m_2}(\mu_H(1),\mu_H(m_2))>0$.
\end{itemize}
Thus, in both cases, such $\mu$ is strictly suboptimal. Hence, we must have $\mu_H^*(1)=1$.


We now show by contradiction that the optimal policy is such that  $\mu_H^*(m)>0,\forall m>1$.
Thus, let $\mu$ be a policy such that $\mu_H(1)= 1$ and 
assume by contradiction that $\exists m_1>1:\mu_H(m_1)=0$. Let $m_2=1$.
Clearly, $m_1\neq m_2$,  $0=\mu_H(m_1)<1$ and $1=\mu_H(m_2)>0$. Therefore, 
we can apply the framework developed in the preliminary part of the proof.
Indeed, we have
\begin{align}
\nonumber
&f_{m_1,m_2}(0,\mu_H(m_2))=f_{m_1,1}(0,1)=1>0.
\end{align}
 Therefore, $\mu$ is strictly suboptimal, hence $\mu_H^*(m){>}0,\forall m{>}1$.
  
 We now show that $\mu_H^*(m)\leq 1/m,\forall m>1$. In fact, from (\ref{R}) and (\ref{P}) we obtain
 \begin{align}
 \nonumber
&\frac{\mathrm d\bar R(\mu)}{\mathrm d\mu_H(m)}=
\left(\begin{array}{c}N\\m\end{array}\right)\pi_H^m\pi_L^{N-m}m\left(1{-}\mu_H(m)\right)^{m-2}
\\&
\qquad\qquad\qquad\times\left(1-m\mu_H(m)\right),
\\
&\frac{\mathrm d\bar Q_H(\mu)}{\mathrm d\mu_H(m)}
=
\left(\begin{array}{c}N-1\\m-1\end{array}\right)\pi_H^{m-1}\pi_L^{N-m}>0.
\end{align}
Note that $\bar Q_H(\mu)$ is an increasing function of $\mu_H(m)$, whereas 
$\bar R(\mu)$ is increasing for $\mu_H(m)<m^{-1}$, decreasing for $\mu_H(m)<m^{-1}$, and achieves the maximum at $\mu_H(m)=m^{-1}$.
Therefore, any  $\mu_H(m)>m^{-1}$ is suboptimal:
by decreasing $\mu_H(m)$, one obtains a strictly larger network reward and strictly smaller average power consumption (which, thus, still satisfies the constraint (\ref{constraints})).
Necessarily, $0<\mu_H^*(m)\leq 1/m,\forall m>1$.

 Thus, let $\mu$ be a policy such that $\mu_H(1)= 1$ and $0<\mu_H(m)\leq m^{-1},\forall m>1$,
 and let $m_1>m_2>1$.
Since $\mu_H(m_1)\in(0,1)$ and $\mu_H(m_2)\in(0,1)$,
in order to be optimal, $\mu$ needs to satisfy $f_{m_1,m_2}(\mu_H(m_1),\mu_H(m_2))=0$, yielding
 \begin{align*}
 & (1-\mu_H(m_1))^{m_1-2}[1-m_1\mu_H(m_1)]
  \\&
  =(1-\mu_H(m_2))^{m_2-2}[1-m_2\mu_H(m_2)],
 \end{align*}
 for all pairs $m_1>m_2>1$,
 and therefore we obtain (\ref{mona2}), repeated here for convenience,
  \begin{align}
  \label{mona}
  \left(1-\mu_H(m)\right)^{m-2}\left(1-m\mu_H(m)\right)=\phi,\ \forall m>1,
\end{align}
where $\phi\in[0,1)$ is a constant (note that, since $\mu_H(m)>0$, necessarily $\phi<1$).
 The left hand side of (\ref{mona}) is a strictly decreasing function of $\mu_H(m)\in (0,m^{-1})$, which equals 
$1$ for $\mu_H(m)=0$ and $0$ for $\mu_H(m)=1/m$. Therefore, there exists a unique $\mu_H(m)\in(0,1/m]$,
denoted as $\mu_H^{(\phi)}(m)$,
which satisfies (\ref{mona}) with equality. 

Thus, it remains to prove that the optimal policy is given by $\mu_H^*(1)=1$, 
$\mu_H^*(m)=\mu_H^{(\phi)}(m),\forall m>1$, for some $\phi\in[0,1)$. 
To conclude, we need to determine such $\phi$.

Since the left hand side of (\ref{mona}) is a strictly decreasing function of $\mu_H(m)\in (0,m^{-1})$,
it follows that $\mu_H^{(\phi)}(m)$  is a strictly decreasing function of  $\phi$,
with $\mu_H^{(0)}(m)=m^{-1}$
 and $\lim_{\phi\to 1}\mu_H^{(\phi)}(m)=0$.
Therefore, for $1>\phi_1>\phi_2\geq 0$, 
we get  $\mu_H^{(\phi_1)}(m)<\mu_H^{(\phi_2)}(m),\forall m>1$, hence, 
by inspection of (\ref{P}), we obtain
\begin{align}
\label{Qincr}
&\bar Q(\mu^{(\phi_1)})<\bar Q(\mu^{(\phi_2)}).
\end{align}
Similarly, since $\bar R(\mu)$ is an increasing function of $\mu_H(m)\in(0,m^{-1}),\forall m>1$,
by inspection of (\ref{R}) we obtain
\begin{align}
\label{Rincr}
&\bar R(\mu^{(\phi_1)})<\bar R(\mu^{(\phi_2)}).
\end{align}
Hence, $\bar R(\mu^{(\phi)})$ and  $\bar Q(\mu^{(\phi)})$ are strictly decreasing functions of  $\phi\in[0,1)$.
When $\phi=0$, we obtain $\mu_H^{(0)}(m)=m^{-1}$, hence
\begin{align}
&\bar Q_H(\mu^{(0)})
=
\sum_{m=0}^{N-1}\left(\begin{array}{c}N-1\\m\end{array}\right)\pi_H^m\pi_L^{N-1-m}\frac{1}{m+1}
\\&
=
\frac{1}{N\pi_H}\sum_{m=1}^{N}\left(\begin{array}{c}N\\m\end{array}\right)\pi_H^{m}\pi_L^{N-m}
\\&
=
 \frac{(\pi_H+\pi_L)^N-\pi_L^{N}}{N\pi_H}
 =
\frac{\lambda_{H,\max}}{P_{tx}},
\end{align}
where we have used the fact that $\pi_H+\pi_L=1$,
and the definition of $\lambda_{H,\max}$.

Therefore, when $\lambda_H\geq\lambda_{H,\max}$, from (\ref{Qincr}) and (\ref{Rincr}), for all $\phi\in[0,1)$ we obtain
\begin{align}
&\bar R(\mu^{(\phi)})\leq\bar R(\mu^{(0)}),
\\
&\bar Q(\mu^{(\phi)})\leq\bar Q(\mu^{(0)})=\frac{\lambda_{H,\max}}{P_{tx}}\leq\frac{\lambda_{H}}{P_{tx}}.
\end{align}
Clearly, $\mu^{(0)}$ satisfies the constraint (\ref{constraints}) and achieves the maximum network reward over $\phi\in[0,1)$.
Hence, the policy $\mu_H^{(0)}(m)=m^{-1}$ is optimal when $\lambda_H\geq\lambda_{H,\max}$, thus proving
(\ref{case3}).

On the other hand, when $\lambda_H<\lambda_{H,\max}$, policy $\mu_H^{(0)}(m)$ violates the constraint (\ref{constraints}).
Necessarily, in this case $0<\phi<1$. Note that, in the limit $\phi\to1$,
we obtain $\mu_H^{(1)}(m)=0,\forall m>1$, hence $\lim_{\phi\to1}\bar Q(\mu^{(\phi)})=\pi_L^{N-1}$.
Thus, $\forall \phi\in(0,1)$ we obtain
\begin{align*}
&\pi_L^{N-1}=\lim_{\phi\to1}\bar Q(\mu^{(\phi)})<\bar Q(\mu^{(\phi)})<\bar Q(\mu^{(0)})=\frac{\lambda_{H,\max}}{P_{tx}},
\end{align*}
and, by assumption,
\begin{align}
&\pi_L^{N-1}<\frac{\lambda_H}{P_{tx}}<\frac{\lambda_{H,\max}}{P_{tx}}.
\end{align}
Therefore, there exists a unique $\hat\phi\in(0,1)$ such that $\bar Q(\mu^{(\hat\phi)})=\frac{\lambda_H}{P_{tx}}$.
Under such $\hat\phi$, we obtain
\begin{align}
\left\{
\begin{array}{l}
\bar R(\mu^{(\phi)})<\bar R(\mu^{(\hat\phi)}),
\\
\bar Q(\mu^{(\phi)})<\bar Q(\mu^{(\hat\phi)})=\frac{\lambda_H}{P_{tx}},
\end{array}
\right.\ \forall 0<\phi<\hat\phi,
\end{align}
and 
\begin{align}
\bar Q(\mu^{(\phi)})>\bar Q(\mu^{(\hat\phi)})=\frac{\lambda_H}{P_{tx}},
\ \forall \hat\phi<\phi<1.
\end{align}
We conclude that any $\phi>\hat\phi$ violates the constraint, whereas
any $\phi<\hat\phi$ is strictly suboptimal. Thus, $\hat\phi$ is the optimal value among $\phi\in(0,1)$ which maximizes the 
network reward under the constraint (\ref{constraints}).

The theorem is thus proved.
\end{proof}

\section{Proof of Theorem \ref{thm2}}
\label{proofofthm2}
\begin{proof}
Assume node $n$ is in the "high" EH state in slot $k$.
Then, the expected transmission probability satisfies
\begin{align}
\label{orignial}
&\mathbb E\left[q_{n,k}|S_{n,k}=H\right]
=
\mathbb P\left(q_{n,k}=1|S_{n,k}=H\right)
\nonumber\\&
=
\sum_{\mathbf t}
\mathbb P\left(q_{n,k}=1,\mathbf t_k=\mathbf t|S_{n,k}=H\right),
\end{align}
where $\mathbf t_k{=}(t_0,t_1,\dots,t_{k-1})\in\{0,1,\dots, N\}^k$
is the vector of nodes that transmit from slot $0$ to slot $k-1$.
Then, using Bayes' rule,
\begin{align}
\label{expr}
&\mathbb P\left(q_{n,k}{=}1,\mathbf t_k{=}\mathbf t|S_{n,k}{=}H\right)
=
\mathbb P\left(q_{n,k}{=}1|\mathbf t_k{=}\mathbf t,S_{n,k}{=}H\right)
\nonumber\\&
\qquad\times
\frac{\mathbb P\left(S_{n,k}=H|\mathbf t_k=\mathbf t\right)\mathbb P\left(\mathbf t_k=\mathbf t\right)}{\mathbb P(S_{n,k}=H)}.
\end{align}
Note that the belief in slot $k$ available at the gateway is a function of $\mathbf t_k$,
as can be seen from (\ref{belief}). 
This can be proved by induction. In fact, 
$\beta_0(m)=\mathbb P(N_{H,0}=m)$ (prior at time $0$),
$\mu_0$ is a function of $\beta_0$ via (\ref{dfghcv}),
and $\beta_1$ is a function of $(\beta_0,\mu_0,t_0)$ via (\ref{betaupdate});
thus, $\beta_1$ is a function of $t_0$.
Then, assuming that $\beta_{k-1}$ is a function of $\mathbf t_{k-1}$,
we have the following: $\mu_{k-1}$ is a function of $\beta_{k-1}$ via (\ref{dfghcv}),
and $\beta_k$ is a function of $(\beta_{k-1},\mu_{k-1},t_{k-1})$ via (\ref{betaupdate});
thus, $\beta_k$ is a function of $\mathbf t_k=(\mathbf t_{k-1},t_{k-1})$,
and by induction $\beta_k$ is a function of $\mathbf t_k$, for all $k\geq 0$. 
We denote this function as $\beta_k(m)=\psi_k(m|\mathbf t_k)$, and we denote (\ref{dfghcv}) computed under such $\mathbf t_k$ as $\mu_k(\mathbf t_k)$.
It follows that 
\begin{align}
\label{expr5}
&\mathbb P\left(q_{n,k}=1|\mathbf t_k=\mathbf t,S_{n,k}=H\right)=\mu_k(\mathbf t)
\nonumber\\&
=
\frac{\sum_{m=1}^N\psi_k(m|\mathbf t)m\mu_H^*(m)}{\sum_{m=1}^N\psi_k(m|\mathbf t)m}.
\end{align}
Additionally,
\begin{align}
\label{expr3}
&\mathbb P\left(S_{n,k}=H|\mathbf t_k=\mathbf t\right)
=
\sum_{m=1}^N\mathbb P\left(S_{n,k}{=}H,N_{H,k}{=}m|\mathbf t_k{=}\mathbf t\right)
\nonumber\\&
\!\!\!\!=\!\!
\sum_{m=1}^N
\mathbb P\left(S_{n,k}{=}H|N_{H,k}{=}m,\mathbf t_k{=}\mathbf t\right)
\mathbb P\left(N_{H,k}{=}m|\mathbf t_k{=}\mathbf t\right),
\end{align}
where we have marginalized with respect to the number of active nodes $N_{H,k}$ (clearly, $N_{H,k}>0$ since $S_{n,k}=H$).
Note that, independently of $\mathbf t_k$, when $N_{H,k}=m$ nodes are active, the probability that $S_{n,k}=H$ is $m/N$. In fact, 
nodes are identical to each other. We thus obtain
\begin{align}
\label{expr4}
&\mathbb P\left(S_{n,k}=H|\mathbf t_k=\mathbf t\right)=
\frac{1}{N}\sum_{m=1}^Nm\mathbb P\left(N_{H,k}{=}m|\mathbf t_k{=}\mathbf t\right)
\nonumber\\&
=\frac{1}{N}\sum_{m=1}^Nm\psi_k(m|\mathbf t),
\end{align}
where the second step follows from the definition of $\psi_k(m|\mathbf t)$.
By replacing (\ref{expr4}) and (\ref{expr5}) into (\ref{expr}), we thus obtain
\begin{align}
\label{exprx}
&\mathbb P\left(q_{n,k}{=}1,\mathbf t_k{=}\mathbf t|S_{n,k}{=}H\right)
\nonumber\\&
=
\frac{1}{N\mathbb P(S_{n,k}=H)}
\sum_{m=1}^N\psi_k(m|\mathbf t)\mathbb P\left(\mathbf t_k{=}\mathbf t\right)
m\mu_H^*(m)
\nonumber\\&
=
\frac{1}{N\mathbb P(S_{n,k}=H)}
\sum_{m=1}^N\mathbb P\left(N_{H,k}{=}m,\mathbf t_k{=}\mathbf t\right)
m\mu_H^*(m),
\end{align}
where in the last step we have used the definition of $\psi_k(m|\mathbf t)$.
Finally, by replacing (\ref{exprx}) into (\ref{orignial}), we obtain
\begin{align}
\label{sdfggfds}
&\mathbb E\left[q_{n,k}|S_{n,k}=H\right]
\nonumber\\&
=
\frac{1}{N\mathbb P(S_{n,k}=H)}\sum_{m=1}^N\mathbb P\left(N_{H,k}{=}m\right)m\mu_H^*(m),
\end{align}
where we have marginalized with respect to $\mathbf t_k$.

On the other hand, with the genie-aided scheme, we obtain
\begin{align}
\label{orignial2}
&\mathbb E\left[q_{n,k}|S_{n,k}=H\right]
=
\mathbb P\left(q_{n,k}=1|S_{n,k}=H\right)
\nonumber\\&
=
\sum_{m=1}^N
\mathbb P\left(q_{n,k}=1|N_{H,k}=m,S_{n,k}=H\right)
\nonumber\\&
\quad\times
\frac{\mathbb P\left(S_{n,k}=H|N_{H,k}=m\right)\mathbb P\left(N_{H,k}=m\right)}{\mathbb P(S_{n,k}=H)},
\end{align}
where we have marginalized with respect to $N_{H,k}$ and used Bayes' rule.
Note that, in the genie-aided scheme, $\mathbb P\left(q_{n,k}=1|N_{H,k}=m,S_{n,k}=H\right)=\mu_H^*(m)$.
Additionally, $\mathbb P\left(S_{n,k}=H|N_{H,k}=m\right)=m/N$, since nodes are identical.
Thus, we finally obtain
 \begin{align}
\label{orignial2}
&\mathbb E\left[q_{n,k}|S_{n,k}=H\right]
\nonumber\\&
=
\frac{1}{N\mathbb P(S_{n,k}=H)}\sum_{m=1}^N\mathbb P\left(N_{H,k}{=}m\right)m\mu_H^*(m),
\end{align}
which is the same expression as (\ref{sdfggfds}) for the Bayesian case.

We conclude that, in every slot, the expected transmission probability (hence the expected power consumption) is the same under the genie-aided and Bayesian schemes.
The theorem is proved.
\end{proof}

\section{Bisection algorithms}
\label{app_algo}
We now present two bisection algorithms, to compute the value of  $\phi\in(0,1)$ in Theorem \ref{thm1},
and to compute the  $\mu_H(m)$ for a given $\phi$ in (\ref{mona2}), respectively.
To this end, note
from the proof of Theorem \ref{thm1} in Appendix \ref{proofofthm1} that
$\bar Q(\mu^{(\phi)})$ and $\bar R(\mu^{(\phi)})$
are strictly decreasing functions of $\phi$, see (\ref{Qincr}) and (\ref{Rincr}).
It follows that,
if $\bar Q_H(\mu^{(\tilde\phi)})<\min\left\{1,\frac{\lambda_H}{P_{tx}}\right\}$, then $\phi>\tilde\phi$
and $\tilde\phi$ is a lower bound to $\phi$; vice versa, if 
$\bar Q_H(\mu^{(\tilde\phi)})>\min\left\{1,\frac{\lambda_H}{P_{tx}}\right\}$, then $\phi<\tilde\phi$ and $\tilde\phi$ is an upper bound to $\phi$.
This observation leads to the following bisection algorithm.
 \begin{algo}\label{alg1}[To determine $\phi$ in Theorem \ref{thm1}]$ $
 \begin{itemize}
 \item \emph{\bf Init:} $\phi_{\min}=0$, $\phi_{\max}=1$; accuracy $\epsilon_\phi\ll 1$;
 \item \emph{\bf Main:} $\tilde\phi:=\frac{\phi_{\min}+\phi_{\max}}{2}$.
 Determine $\tilde\mu_H$ via Algorithm \ref{alg2};
if $\bar Q_H(\tilde\mu_H){<}\min\left\{1,\frac{\lambda_H}{P_{tx}}\right\}$, set $\phi_{\min}{:=}\tilde\phi$; otherwise,
set $\phi_{\max}{:=}\tilde\phi$;
\item \emph{\bf Test:} repeat \emph{\bf Main} until 
$\phi_{\max}-\phi_{\min}<\epsilon_\phi$; finally, return
$\phi=\frac{\phi_{\min}+\phi_{\max}}{2}$.
 \end{itemize}
 \end{algo}

 For a given $\tilde\phi$ in Algorithm \ref{alg1},
 we leverage the fact that the right hand expression of
 (\ref{mona2}) is a decreasing function of $\mu_H(m)\in(0,1/m)$.
  This observation leads to the following bisection algorithm.
  \begin{algo}\label{alg2}[To find $\tilde\mu_H$ given $\tilde\phi$]
$\forall m\in\{2,3,\dots, N\}$:
 \begin{itemize}
 \item \emph{\bf Init:} $\tilde\phi$ given; $u_{\min}{=}0$, $u_{\max}{=}1/m$; accuracy $\epsilon_\mu{\ll}1$;
 \item \emph{\bf Main:} $\tilde u{:=}\frac{u_{\min}+u_{\max}}{2}$.
If $(1{-}\tilde u)^{m-2}(1{-}m\tilde u){>}\tilde\phi$,
set $u_{\min}:=\tilde u$; otherwise,
set $u_{\max}:=\tilde us$;
\item \emph{\bf Test:} repeat \emph{\bf Main} until 
$u_{\max}-u_{\min}<\epsilon_\mu$; finally, return
$\tilde\mu_H(m)=\frac{u_{\min}+u_{\max}}{2}$.
 \end{itemize}
 \end{algo}
}{
}

\bibliographystyle{IEEEtran}
\bibliography{IEEEabrv,biblio} 

\end{document}